\documentclass[showpacs,amsmath,amssymb,floatfix,twocolumn]{revtex4}

\usepackage{epsfig}

\begin{document}
 
\sloppy
 
\title{First principles electronic structure of spinel LiCr$_2$O$_4$: 
A possible half-metal ?}

\author{Markus Lauer}
\email{mala@lusi.uni-sb.de}
\affiliation{Fachrichtung Theoretische Physik, 
Universit\"at des Saarlandes,
Postfach 15 11 50, D-66041  Saarbr\"ucken, Germany}
\author{Roser Valent\'\i}
\email{valenti@lusi.uni-sb.de}
\affiliation{Institut f\"ur Theoretische Physik,
Universit\"at Frankfurt, Robert-Mayer-Strasse 8, D-60054 Frankfurt, Germany}
\author{H. C. Kandpal}
\email{kandpal@mail.uni-mainz.de}
\affiliation{Institut f\"ur Anorganische Chemie, Universit\"at Mainz,
Duesberg Weg 10-14, D-55099  Mainz, Germany}

\author{Ram Seshadri}
\email{seshadri@mrl.ucsb.edu}
\affiliation{Materials Department, University of California, Santa Barbara, 
CA 93106 USA}

\begin{abstract}
We have employed first-principles electronic structure calculations to 
examine the hypothetical (but plausible) oxide spinel, LiCr$_2$O$_4$ with
the $d^{2.5}$ electronic configuration. The cell (cubic) and internal (oxygen 
position) structural parameters have been obtained for this compound through 
structural relaxation in the first-principles framework. Within the 
one-electron band picture, we find that LiCr$_2$O$_4$ is magnetic, and a 
candidate half-metal. The electronic structure is substantially different from 
the closely related and well known rutile half-metal CrO$_2$. In particular, 
we find a smaller conduction band width in the spinel compound, perhaps as a 
result of the distinct topology of the spinel crystal structure, and the 
reduced oxidation state. The magnetism and half-metallicity of LiCr$_2$O$_4$ 
has been mapped in the parameter space of its cubic crystal structure. 
Comparisons with superconducting LiTi$_2$O$_4$ ($d^{0.5}$), heavy-fermion 
LiV$_2$O$_4$ ($d^{1.5}$) and charge-ordering LiMn$_2$O$_4$ ($d^{3.5}$) suggest 
the effectiveness of a nearly-rigid band picture involving simple shifts of 
the position of $E_{\rm F}$ in these very different materials. Comparisons are 
also made with the electronic structure of ZnV$_2$O$_4$ ($d^{2}$), a 
correlated insulator that undergoes a structural and antiferromagnetic phase 
transition.

\end{abstract}

\pacs{71.15.Nc, 
      71.20.-b, 
      75.50.-y  
              }

\maketitle

\section{Introduction}

The stoichiometric, self-doped oxide spinels LiM$_2$O$_4$ 
(with M = Ti, V and Mn) can be said to constitute a periodic ``hall of fame''.
In all of them, Li is monovalent and the oxygen ion, negative
and divalent; this means that the transition metal oxidation state is
formally 3.5+. The number of $d$ electrons per transition metal atom would
therefore be 0.5, 1.5, 2.5 or 3.5 when M = Ti, V, Cr or Mn respectively.
LiTi$_2$O$_4$ is a superconductor with a $T_C$ of 12 K. When 
first prepared in 1973 \cite{liti2o4} it was described as a high temperature 
superconductor. Along with BaBi$_{1-x}$Pb$_x$O$_3$ \cite{sleight},  it was 
amongst the only oxide superconductors with $T_C > 10$ K until the advent of 
the layered cuprates. It is also the \textit{only} metallic oxide spinel 
known. Its neighbor, LiV$_2$O$_4$ is already on the verge of localization, 
and as a result, displays heavy-fermion behavior \cite{liv2o4} --- one of 
the few systems without $f$ electrons that is known to do so --- LiMn$_2$O$_4$,
one more member from this family, is not only a very important cathode 
material in secondary Li-ion batteries \cite{battery},  but also displays 
charge-ordering \cite{rodriguez,basu1},  associated with half-integral charge 
that can order over two distinct sites. Indeed, the well-known Verweij 
transition in ferrite associated with the ordering of charge on the spinel
lattice \cite{verweij} can perhaps be best understood by examining 
LiMn$_2$O$_4$. Near the charge-ordering temperature of this compound, a 
magnetic field can influence electrical transport and this translates to
negative magnetoresistence \cite{basu2}. 

Conspicuous in this list by its absence is the spinel LiCr$_2$O$_4$. Attempts
to prepare this compound are fraught with the difficulty of stabilizing Cr 
in a formal oxidation state that is somewhere between 3+ and 4+. Indeed 
ferromagnetic CrO$_2$ with the rutile structure is well known to require high 
oxygen partial pressures for its preparation. CrO$_2$ decomposes at ambient 
oxygen partial pressures when the temperature exceeds 473 K 
\cite{cro2_prepn,kaemper}.  Given that (i) the isostructural compounds 
neighboring LiCr$_2$O$_4$  are so interesting, and (ii) the structure and oxidation state
 of  LiCr$_2$O$_4$ are reminiscent of CrO$_2$, the prototypic oxide half-metal 
\cite{kaemper,cro2_point}, we thought the compound LiCr$_2$O$_4$ worthy of study.

In this work, we use spin-polarized first principles density functional 
calculations within two different schemes: Full potential linearized augmented
plane-wave calculations \cite{WIEN97,WIEN2k} that permit an optimization of the 
geometry of the cubic spinel compound (lattice parameter and internal 
structural parameter) and Linear Muffin Tin Orbital
(LMTO) 
 calculations \cite{andersen} which 
are geared to the visualization of spin-resolved bonding. The electronic 
structure of LiCr$_2$O$_4$ is compared with that of rutile CrO$_2$, 
LiM$_2$O$_4$ (M = Ti, V and Mn), and ZnV$_2$O$_4$. The parameter space of 
half-metallicity in LiCr$_2$O$_4$ is mapped out in terms of the two structural 
parameters. 

\section{Crystal structures and computational methods}

\begin{figure}[thb]
\epsfig{file=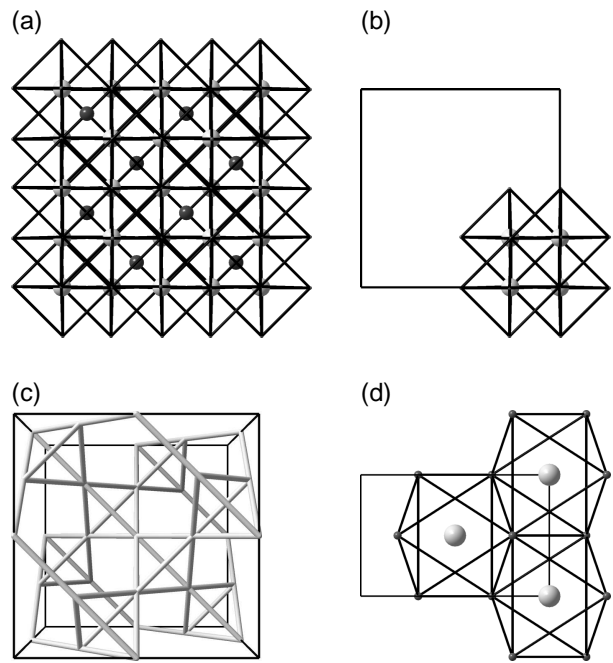, width=8cm}
\caption{(a) Spinel AB$_2$O$_4$ structure showing octahedral B 
atoms (grey spheres) at the centers of O$_6$ octahedra (depicted using 
sticks). The black spheres are the tetrahedrally coordinated A atoms. (b) 
Four BO$_6$ octahedra in the spinel structure showing the formation of B$_4$
tetrahedra as well as the nature of the edge-sharing of oxygen. (c) Network of 
corner-sharing B$_4$ tetrahedra in the spinel structure. (d) Portion of the
crystal structure of rutile CrO$_2$ showing Cr atoms (grey spheres) in the
middle of O$_6$ octahedra.}
\label{fig01}
\end{figure}

The cubic spinel crystal structure AB$_2$O$_4$
 is completely described by two parameters,
the cubic cell parameter $a$, and the internal oxygen positional parameter 
$(x,x,x)$ where $x \sim 0.25$ \cite{spinel_struc}. The A atoms (Li) occupy 
tetrahedral sites created by a network of BO$_6$ octahedra where B is here a
transition metal. The octahedra of oxygen around B are perfectly regular 
when $x=0.25$. Deviations $\delta$, $x=0.25 + \delta$ indicate either a 
trigonal expansion ($\delta > 0$) or compression ($\delta < 0$), achieved by 
pushing two opposite triangular faces of the octahedron apart or together.
The B atom sublattice (Cr in LiCr$_2$O$_4$) comprises a network of B$_4$ 
tetrahedra. Because the B atoms form tetrahedra, antiferromagnetic ordering 
is frustrated on the B site of the spinel \cite{anderson} in a manner 
reminiscent of the residual entropy problem in the crystal structure of 
ice I\textit{h} \cite{pauling}. Various views of the spinel crystal structure,
and the structure of rutile CrO$_2$ are shown in Fig.\ \ref{fig01}.

Scalar-relativistic Kohn-Sham equations were solved taking all relativistic
effects into account except for the spin-orbit coupling. We performed both
full potential linearized augmented plane-wave (LAPW)
 calculations based on the \textsc{WIEN97} and \textsc{WIEN2k} codes 
\cite{WIEN97,WIEN2k} as well as Linear muffin-tin orbital (LMTO) calculations 
\cite{andersen} within the atomic sphere approximation (ASA) where we used the 
\textsc{stuttgart tb-lmto-asa} program \cite{stuttgartLMTO}. The generalized
gradient approximation (GGA)  for the calculation of exchange correlation was 
considered following the Perdew-Burke-Ernzerhof \cite{Perdew_96} 
parameterization for the LAPW calculations and the Perdew-Wang prescription
\cite{perdew-wang} for the LMTO calculation\cite{comment_GGA}. 

In our LAPW calculations we considered twenty inequivalent sampling $k$ points 
and the modified tetrahedron method \cite{Bloechl_94} for the Brillouin zone 
integration during the self-consistent iterations. In order to test the accuracy
of the sampling,
 we also performed calculations with up to 120 and 195 irreducible $k$-points
 without observing significant qualitative changes in the results apart that some
spikes in the density of states plots get rounded.  We set the 
energy threshold between core and valence states at -6 Ryd
\cite{comment_2}.
 For the number of 
plane waves, the criterion used was RMT (Muffin-Tin Radius) $\times$ 
$k_{\rm max}$ (for the plane waves) = 8.  We considered various sets of 
muffin-tin radii to ensure well converged spin-polarized calculations.

In our LMTO calculations 256 irreducible $k$ points were used in the primitive 
wedge of the BZ.  Space filling in the unit cell of LiCr$_2$O$_4$ was ensured
through the use of two empty spheres  with bases of $s$ and $p$ orbitals
 at (0, 0, 0)  and ($y$, 1/8,
1/8) where the value of $y$ is set automatically by the code according
to the oxygen position. 
LMTO electronic structures were analyzed by calculating crystal orbital 
Hamiltonian populations (COHPs) \cite{cohp} which are densities of states 
weighted by appropriate Hamiltonian matrix elements. COHPs are indicative of 
the strength and nature of a bonding (positive COHP) or antibonding (negative 
COHP) interaction. The signs we use here are the opposite of what is used in 
the original definition of Dronskowski and Bl\"ochl \cite{cohp}. LMTO-ASA 
calculations do not allow for structural relaxation while structural
relaxation process is implemented in the LAPW package. We will present in
what follows a comparison of the results obtained by both methods.

\section{Crystal and electronic structure of $\text{LiCr}_2\text{O}_4$}

\subsection{Structure optimization}

\begin{figure}[thb]
\epsfig{file=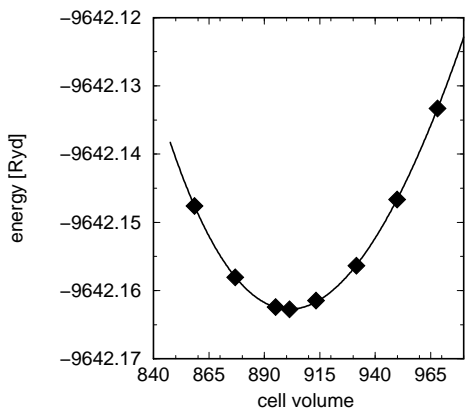, width=7cm}
\caption{LAPW energy versus primitive unit cell volume (in Bohr$^3$) for the 
LiCr$_2$O$_4$ system. The squares are calculated LAPW data and the
 dashed curve
 is a fit 
to the Murnaghan equation of state \protect\cite{Murnaghan_44}}
\label{fig02}
\end{figure}

Within the LAPW scheme we performed a structure optimization for 
LiCr$_2$O$_4$ where the cubic cell parameter $a$ and the oxygen position 
$x$ were varied till the optimal structure was obtained. In order to 
perform the optimization we considered two steps: (i) We first took several 
crystal structures with different lattice constants $a$ and the same $x$.  
For each of them we performed an iteration procedure up to self-consistency 
and we compared the total energies.  The energy minimum then defines the
optimal $a$. For LiCr$_2$O$_4$ the energy minimum was reached at 
$a$ = 8.11(4) \AA\  (see Fig.\ \ref{fig02}).

 (ii) In a second step we proceeded to optimize 
the oxygen positions by evaluating the atomic forces 
\cite{Kohler_96}. The equilibrium oxygen position was obtained using a damped 
Newton dynamics method according to the expression

\begin{eqnarray}
R^{t+1}_m = R^t_m + \eta_m (R_m^t - R_m^{t-1}) + \delta_m F_m^t
\end{eqnarray}

\noindent
where $R_m^t$ and $F_m^t$ are the coordinate and force for the atom $m$ at 
time step $t$. $\delta_m$ determines the speed of motion and $\eta_m$ changes 
from $\eta_m$ to $1 - \eta_m$ if the force changes its direction from one step 
to the next.  
The optimal value we obtained for  the oxygen position was $x$= 0.253(5)
which implies a small distortion from the perfect spinel structure. A further
unit-cell optimization did not change the result significantly.

In the previous calculation scheme the unit-cell is first optimized and then
the structure relaxed. Since the two operations do not ne\-cessa\-rily commute, 
we considered also first a structural relaxation in a bigger unit cell 
followed by a lattice cons\-tant optimization. The two procedures do not
yield exactly the same optimal value for $a$ and $x$, but deviations
from  $a = 8.11(4)$ and $x = 0.253(5)$  are small and can be considered within 
the error bars.

We checked both in LAPW and LMTO that the ferromagnetic spin-polarized configuration is lower in 
energy than the non-spin-polarized case. 
 We also considered within LAPW an antiferromagnetic spin arrangement of the
 Cr
atoms in the spinel lattice which showed to be energetically
unfavourable with respect to the ferromagnetic arrangement.

\begin{figure}[t]
\epsfig{file=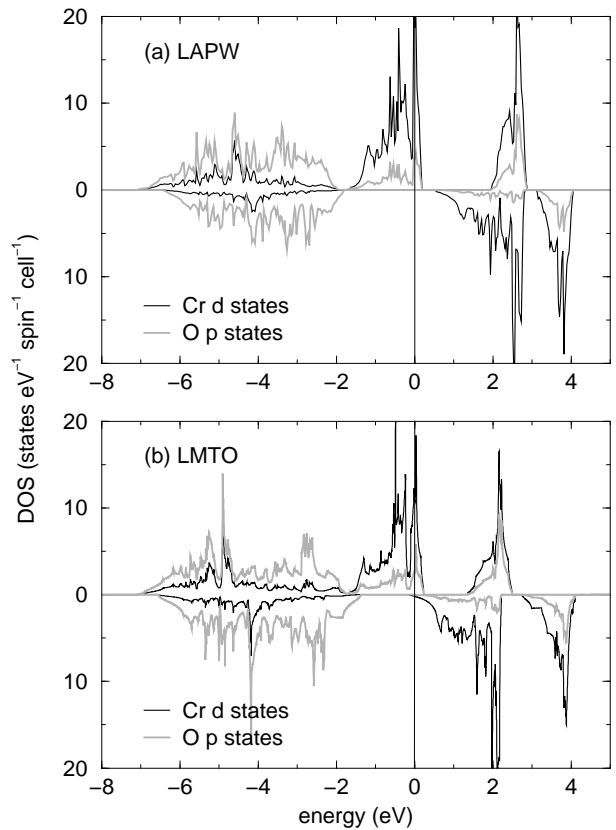, width=8cm}\\
\caption{(a) LAPW and (b) LMTO projected densities of states for 
LiCr$_2$O$_4$. Black traces indicate Cr $d$ states and grey traces indicate
O $p$ states. Spin up states are shown in the upper portions and spin down 
states in the lower portions in both panels. The origin on the energy axis 
in this and in the following plots is the $E_{\rm F}$.} 
\label{fig03}
\end{figure}

\subsection{Density of states (DOS)}

Figure \ref{fig03} compares the partial density  
 of chromium $d$ states and oxygen $p$ 
states from spin polarized (a) LAPW calculations and (b) LMTO calculations.
In both plots, the Fermi energy is taken as the origin on the energy axis.
The upper panel of each plot displays majority (up) spin states and the lower 
panel minority (down) spin states. Cr-$d$ states are shown using black traces 
and O-$p$ state using grey traces. The optimized crystal structure obtained 
from the LAPW calculations was used as the input crystal structure for the
(unrelaxed) LMTO calculations. We note immediately that in terms of gross
features, the two computational methods yield quite similar results, in terms 
both of band positions and widths, as well as details in the structures of 
the projected densities of state. Differences in the two plots arise only as
a result of slightly larger gaps between occupied and unoccupied states 
between the two computational schemes; the LAPW calculations indicate 
 larger (by 0.5 eV) crystal field splitting as well as larger exchange
splitting (by the same magnitude).  This different energy shift can
be understood in terms of the different nature of the two schemes and
 that for open structures -as the spinels are- the LMTO approach is not as
accurate as the LAPW approach. We use here therefore the LAPW results as a reference for accuracy.

O $p$ states are found to be largely centered around -4.5 eV and possess
a width of about 5 eV. There are also Cr $d$ states in this energy region,
indicative of covalency. Near $E_{\rm F}$ we find majority Cr $t_{2g}$ ($\uparrow$)
 and no minority spins in the LAPW DOS, i.e. the polarization is 100\%
 what
characterizes the system as half-metallic.  In the LMTO DOS there is a
residual
spin density  of minority Cr $t_{2g}$ ($\downarrow$) spins at  $E_{\rm
F}$.  While strictly speaking the LMTO results predict an {\it
approximate}
half-metal behavior, the small  overall differences between both
LMTO and LAPW approaches justifies the use of the LMTO-ASA calculations
in some of the subsequent discussion.

 The $t_{2g}(\uparrow)$ states have a width of about
2 eV and are bimodal, with the two modes separated by a pseudo-gap. The 
$E_{\rm F}$ lies in the center of a peak in the DOS. The crystal field
splitting between $t_{2g}(\uparrow)$ and $e_g(\uparrow)$ is of the order
of 3 eV (taking the midpoints of the different manifolds) and the 
exchange splitting between $t_{2g}(\uparrow)$ and $t_{2g}(\downarrow)$ 
is of the order of 2 eV. These values are in the LMTO DOS
-as discussed above- about 0.5 eV larger. 

The two-peaked $t_{2g}(\uparrow)$ DOS is puzzling. One suspect is the trigonal 
distortion of the oxygen octahedron around the Cr atoms. In LiCr$_2$O$_4$, 
the calculated oxygen position 0.253(5) suggests a very small distortion from 
perfect octahedral coordination.  We have performed LMTO calculations 
using the ideal oxygen position $x$ = 0.250 corresponding to perfect 
CrO$_6$ octahedra. The differences in the DOS (between the structures where
$x$ = 0.25 and $x$ = 0.253) are negligible and the two-peaked structure of 
the $t_{2g}(\uparrow)$ DOS is retained, and $E_{\rm F}$ continues to fall
in the second DOS peak even when O atoms are in the ideal position and the 
CrO$_6$ octahedra are undistorted.

Bimodal $t_{2g}(\uparrow)$ states, also observed in the electronic structure 
of spinel LiTi$_2$O$_4$ \cite{satpathy},  could 
arise from the restricted B-O-B dispersion in the spinel structure; this 
network can be built up from B$_4$O$_4$ cubes that are interconnected through 
B atoms. The B-O-B bond angles are therefore 90$^{\circ}$. This limits 
dispersion significantly.  In fact, through the analysis of the COHP
 in LiCr$_2$O$_4$ in the next section we can identify that the upper lobe of the DOS
corresponds mainly to Cr-O interactions and  less to Cr-Cr interactions.

 The most important conclusion that we draw from the 
DOS of \textit{ferromagnetic\/} LiCr$_2$O$_4$ is that states at and near 
$E_{\rm F}$ are strongly spin-polarized and the system is indeed a magnetic 
half-metal. However, the fact that $E_{\rm F}$ falls on a sharply peaked
region of the DOS suggests (i) that the compound might be unstable/difficult 
to prepare and (ii) if it is prepared, might be subject to 
electronic 
instabilities associated with electron correlation, i.e.  
the opening of a Mott-Hubbard gap, or charge-ordering.
    We suggest that various different scenarios could stabilize the
structure of LiCr$_2$O$_4$,
  for instance the presence of a slightly distorted antiferromagnetic structure as it is
the case in ZnV$_2$O$_4$.  Also if the system gets a little oxidized,
by removing
for instance some Li to form Li$_{\delta}$Cr$_2$O$_4$, where $\delta <
1$, it may be then possible to have E$_{\rm F}$ sitting in the pseudogap
of the DOS, i.e. at the center of the bimodal DOS and therefore to have
a stable structure.  A third possibility, as mentioned above, is 
 the explicit
consideration of the electron correlation which may open a Mott-Hubbard
gap at E$_{\rm F}$ and therefore settle stability.
\subsection{Comparisons with CrO$_2$}

\begin{figure}[thb]
\epsfig{file=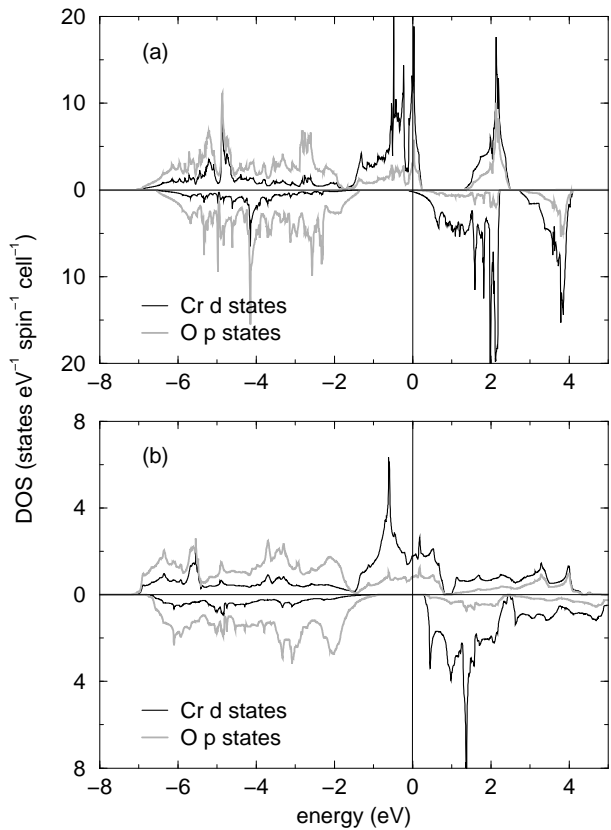, width=8cm}\\
\caption{LMTO projected densities of states for (a) LiCr$_2$O$_4$ and (b)
CrO$_2$. Black traces indicate Cr $d$ states and grey traces indicate
O $p$ states.}
\label{fig04}
\end{figure}

It is important to compare the electronic structure of LiCr$_2$O$_4$ with the 
prototypic oxide half metal CrO$_2$. Present understanding of this material has 
been reviewed recently \cite{coey}. This rutile compound Fig.\ \ref{fig01}(d)) has 
$d^2$ Cr$^{4+}$ surrounded octahedrally by O$^{2-}$. The octahedra form ribbons that 
share edges, and each Cr can therefore be said to bond with two others through 
the shared edge with a Cr-Cr distance of 2.92 \AA ~ \cite{burdett}. 
A number of authors \cite{cro2_calc} have described the electronic structure 
of this material, including Korotin \textit{et al.\/} \cite{korotin} who 
performed LDA + $U$ calculations and argue thereof that the one-electron 
description of a ferromagnetic band metal does not suffice and that 
$U$ = 3.0 eV is required to explain the essential features of this
compound.
Mazin {\it et al.} \cite{Mazin_99} argued  that when the local interaction  is
 small compared to the bandwidth (W)  U/W$\approx$ 0.5 as in CrO$_2$,
 it is disputable which
 method spin-polarized 
LDA or LDA+U provides a better description of the system.

\begin{figure*}[thb]
\epsfig{file=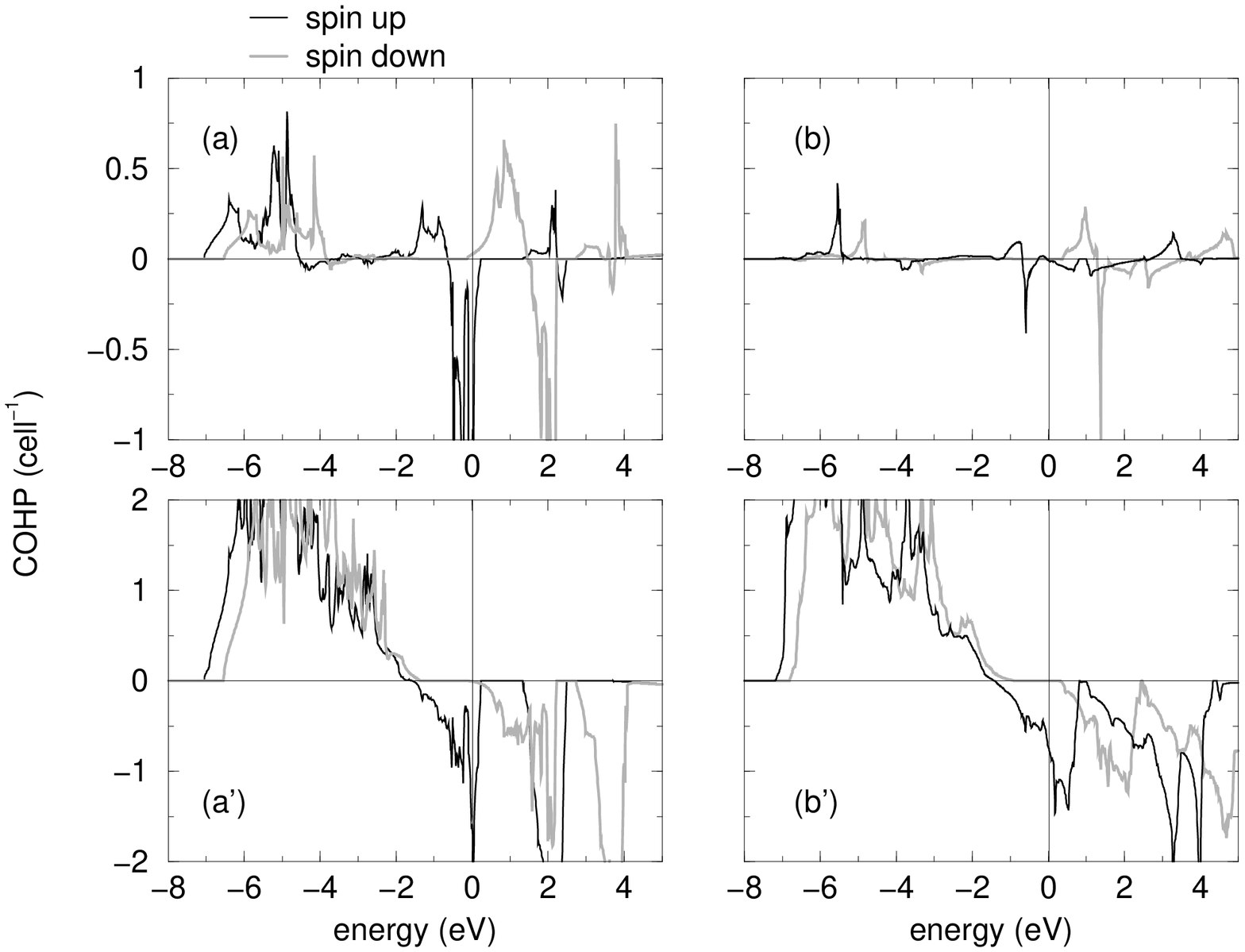, angle= -90, width=16cm}\\
\caption{Crystal orbital Hamiltonian populations (COHPs) in  
LiCr$_2$O$_4$ and CrO$_2$ \textit{per\/} primitive unit cell. 
(a) Cr-Cr interactions in LiCr$_2$O$_4$, (a') Cr-O interactions in 
LiCr$_2$O$_4$,(b) Cr-Cr interactions in CrO$_2$, (b') Cr-O interactions in
CrO$_2$. In each panel, the black trace corresponds to interactions in the
spin up channel and the grey trace to interactions in the spin down channel.}
\label{fig05}
\end{figure*}

We have performed spin-polarized LDA calculations in the LMTO scheme 
 on CrO$_2$ for two reasons. The first is 
that it provides us a consistent comparison with LiCr$_2$O$_4$. The second 
is that spin-polarized bonding through the use of COHPs has not been presented 
before for this important compound. Figure \ref{fig04} compares the Cr $d$ and O $p$ 
DOS of (a) LiCr$_2$O$_4$ and (b) CrO$_2$. O $p$ states of CrO$_2$ are found to 
be quite similar to what is seen for LiCr$_2$O$_4$, spreading from around 
-7 eV to -2 eV. Once again, there is a strong admixture of Cr $d$ states in 
this energy region. The region near the Fermi energy comprised largely of 
Cr $d$ states. The crystal field splitting between $t_{2g}(\uparrow)$ and 
$e_{g}(\uparrow)$ is also similar in both compounds, as might be expected
considering the similarity in charge and coordination. What is significantly
different is the larger $t_{2g}(\uparrow)$ bandwidth in CrO$_2$, as a result
of the different Cr-O topology in the rutile structure, that allows for
easier metal-oxygen-metal hopping. Unlike what we see in LiCr$_2$O$_4$,
the $E_{\rm F}$ in CrO$_2$ does not lay on a peak but instead, lies on the
edge of a pseudo-gap in the $t_{2g} (\uparrow)$ states. This gap arises due to 
a two-in, four-out Jahn-Teller distortion of the CrO$_6$ octahedra 
\cite{burdett}.

Crystal orbital Hamiltonian populations (COHPs) permit a better understanding 
of the precise nature of the different states that would determine the 
properties of LiCr$_2$O$_4$. In Fig.\ \ref{fig05}, we plot (a) Cr-Cr COHPs and 
(a') Cr-O COHPs as a function of energy for LiCr$_2$O$_4$, per primitive unit 
cell. This means Cr interacts with six O (at a distance of 2.00 \AA)
and to six other Cr (at a distance of 2.86 \AA). The COHPs
are spin resolved into interactions between orbitals with majority spin,
and orbitals with minority spin. In the absence of spin-orbit coupling, 
majority states cannot bond with minority states. In the region of the DOS
which was predominantly O-$p$ (-7 eV to -2 eV) we find evidence for quite 
strong Cr-O covalency. At the $E_{\rm F}$, we find a combination of majority 
antibonding Cr-Cr interactions and majority antibonding Cr-O interactions, 
being the Cr-O interaction predominant. There are no minority interactions near the 
$E_{\rm F}$ suggesting that the bonding is highly spin-polarized. This is what 
makes this compound a putative half-metal.

Similar COHPs for CrO$_2$ [Fig.\ \ref{fig05}(b) and (b')] are depicted \textit{per\/}
unit cell. In CrO$_2$, each Cr has 6 oxygen neighbors (4 at 1.90 \AA\/ and 
2 at 1.91 \AA), but only two Cr neighbors (at 2.92 \AA). The shorter Cr-O
distance in CrO$_2$, compared with LiCr$_2$O$_4$ reflects the slightly higher 
oxidation state of Cr. The Cr-O COHP suggests strong Cr-O bonding in both spin 
directions in the region where O $p$ states are found. Cr-Cr bonding is 
insignificant in this structure, as a result of the larger Cr-Cr distance than 
what is found in the spinel, as well as fewer neighbors. This means that 
Cr $t_{2g} (\uparrow)$ states at the $E_{\rm F}$ are largely non-bonding.
As in LiCr$_2$O$_4$, the Cr-O antibonding states at $E_{\rm F}$ are completely 
spin-polarized  and correspond to interactions in the majority spin channel.

\subsection{Comparisons with other LiM$_2$O$_4$, M = Ti, V and Mn, and 
with ZnV$_2$O$_4$}

\begin{figure*}[thb]
\epsfig{file=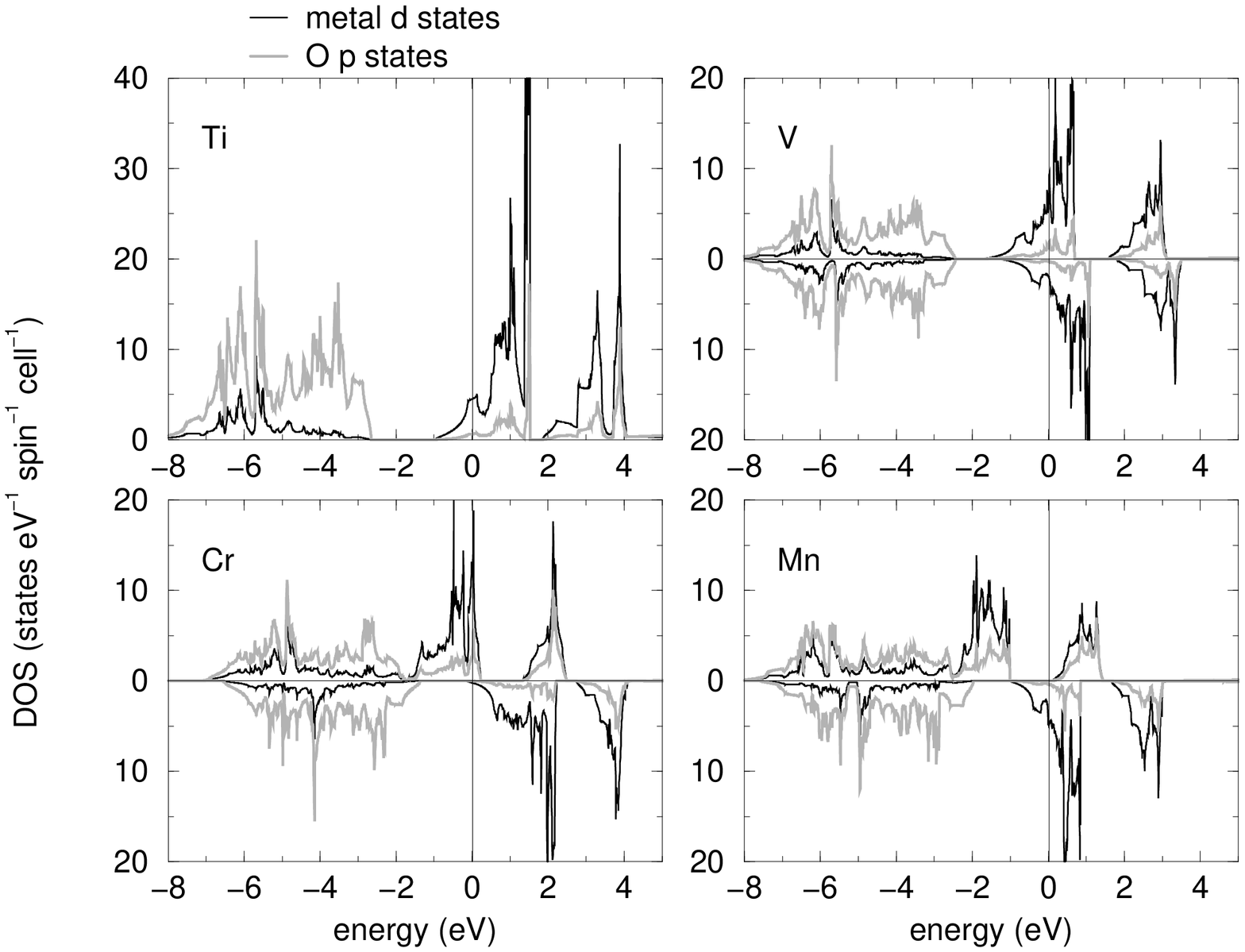, angle = -90, width=16cm}
\caption{LMTO (partial) density of states for LiM$_2$O$_4$ (M= Ti, V, Cr,
 Mn).  Black traces indicate M $d$ states and grey traces indicate
O $p$ states. Both LAPW and LMTO calculations yield spin-polarized ground
states for the compounds with M = V, Cr and Mn.}
\label{fig06}
\end{figure*}

It is of interest to compare the DOS of the compounds LiM$_2$O$_4$, M = Ti, V, 
Cr and Mn. LiTi$_2$O$_4$ is a superconductor. LiV$_2$O$_4$ is a metal on the 
verge of a metal-insulator transition, with correspondingly, an unusually large 
effective carrier mass \cite{liv2o4}.  LiMn$_2$O$_4$ is a correlated 
antiferromagnetic insulator at low temperatures and displays a charge-ordering 
transition near the room temperature \cite{rodriguez,basu1,basu2}.
A number of recent papers describe the electronic structure of LiV$_2$O$_4$
\cite{liv2o4_elec_struc}.  For example, Singh \textit{et al.\/} \cite{singh} 
have performed LAPW calculations on LiV$_2$O$_4$. They find some evidence
for a separation of the $t_{2g}$ mani\-fold into $a_{1g}$ and $e'_g$ states
as a result of the trigonal distortion of the VO$_6$ octahedra, which
is large in this compound. This results in the presence of both flat and
disperse bands near the $E_{\rm F}$, the flat bands presumably giving rise 
to the effective mass enhancement. Mishra and Ceder \cite{mishra} have
reported density functional calculations on LiMn$_2$O$_4$, but with an emphasis on
structural stability, rather than on magnetism.

Figure \ref{fig06}  compares the LMTO DOS of LiM$_2$O$_4$, M = Ti, V, Cr, Mn. 
For the V, Cr and Mn compounds,  we found that the ferromagnetic
spin-polarized
configuration was energetically more favourable that the
non-spin-polarized
configuration.  For the Ti  compound, the
         ferromagnetic solution lies in LAPW and LMTO higher in energy
         than the non-spin-polarized solution.  In fact in LMTO we
          started for this system with a ferromagnetic spin-polarized
         configuration and in the self-consistency cycles it gets
         increasingly non-spin-polarized.
 The width of 
the O $p$ states in all three compounds
is similar. The separation between O $p$ and metal $d$ states is largest in 
the V compound and smallest in the Mn compound. This is in keeping with the
expectation that for a given oxidation state and coordination, moving to the
right amongst transition metals results in  a stabilization of metal 
$d$ states. Such stabilization is in fact the basis of the evolution of 
the band gap in correlated transition metal compounds; from the Hubbard $U$ 
type (arising from $d-d$ Coulomb correlation) for the early transition
metals to charge-transfer $\Delta$ (arising from ligand to metal charge 
transfer) in the later transition metals \cite{zaanen,arima}. The width and 
gross features of metal $t_{2g}$ states in these compounds display many
similarities, and in fact, a simple rigid band picture of the DOS would perhaps
suffice to describe the evolution on going from Ti through Mn. The $t_{2g}$ 
manifold possesses a peaked region at the highest energy in all four
compounds. In the Cr compound, this peaked region coincides with the Fermi 
energy. Interestingly, LiMn$_2$O$_4$ seems to be a low-spin system, with the 
electronic configuration $t_{2g}^3(\uparrow)$, $t_{2g}^{0.5}(\downarrow)$, 
which is distinctly different from the manganese oxides perovskites that 
display  colossal magnetoresistance \cite{pickett,felser}.  In the perovskites,
there is a gap between filled $t_{2g}^3(\uparrow)$ states and the partially 
filled states which are mostly $e_g(\uparrow)$ with some small admixture of 
O $p$ and $t_{2g}(\downarrow)$.
   
\begin{figure}[thb]
\epsfig{file=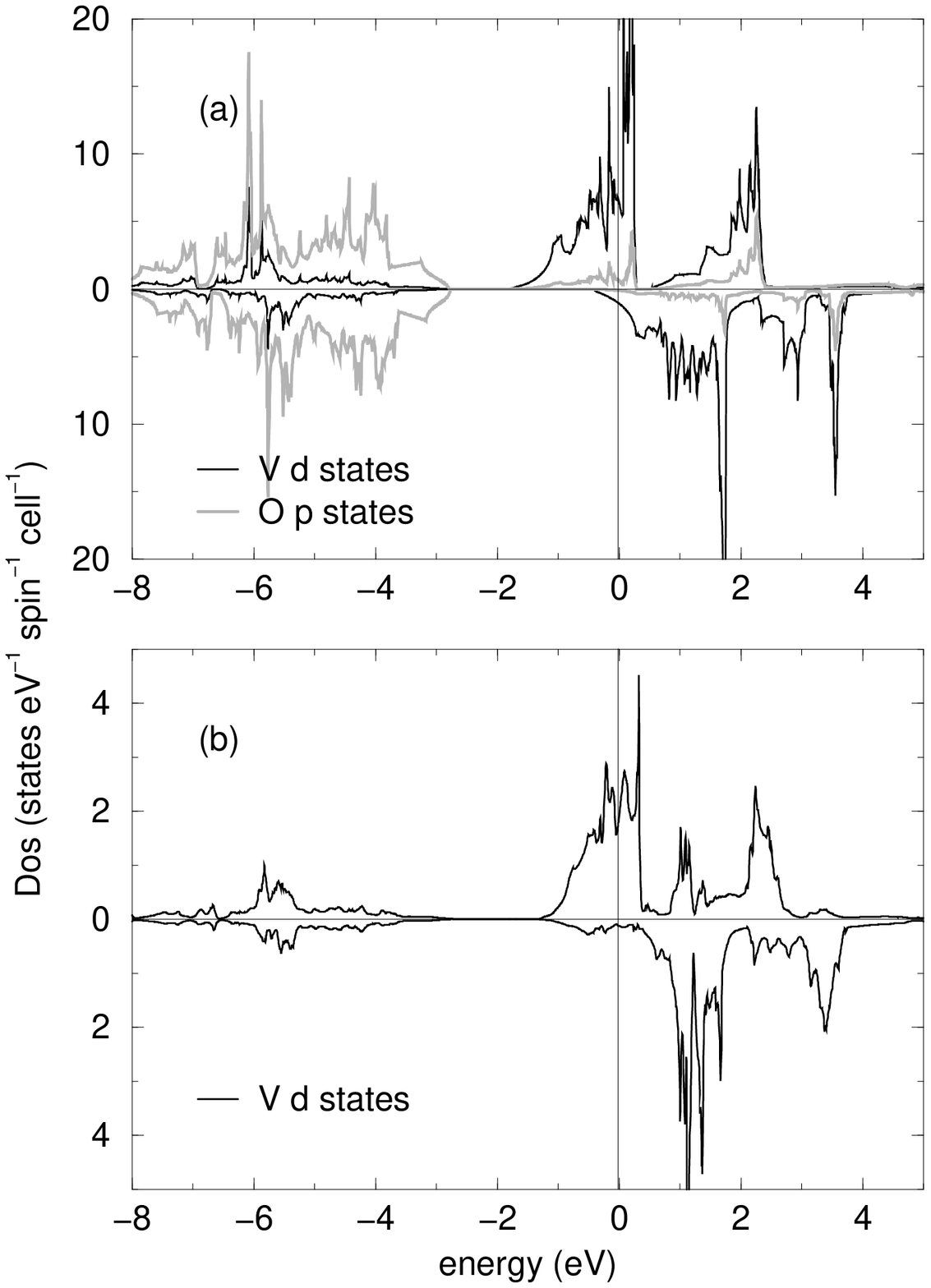, width=8cm}
\caption{(a) Partial DOS for cubic ZnV$_2$O$_4$. Black traces indicate V $d$ 
states and grey traces indicate O $p$ states. (b) V $d$ states for a 
single V atom in antiferromagnetic, tetragonal ZnV$_2$O$_4$.}
\label{fig07}
\end{figure}

The compound ZnV$_2$O$_4$ which is isostructural (spinel) and nearly 
isoelectronic ($d$ electron count of 2 on the B site) with LiCr$_2$O$_4$ 
is known to be a Mott insulator \cite{fujimori} ordering antiferromagnetically 
at 45 K \cite{ueda}, associated with a tetragonal distortion of the spinel
unit cell. Recently Reehuis \textit{et al.\/} \cite{reehuis} have performed
a careful neutron diffraction study on this system. They find a structural
transition from cubic $Fd\overline{3}m$ to tetragonal $I4_1/amd$ at 51(1) K
which breaks the magnetic frustration of the spinel structure and allows an
antiferromagnetic ground state to be obtained (with $T_{\rm N}$ = 40(2) K).
These authors have obtained precise nuclear and magnetic structures at 
60 K (cubic, paramagnetic) and 1.8 K (tetragonal, antiferromagnetic).
We have performed LAPW calculations on the cubic phase, and 
LMTO calculations on using  both the cubic as well as the low temperature
tetragonal structure \cite{reehuis}. The LAPW structure optimization gave
a cell parameter of 8.348(3) \AA\/ to be compared with the 60 K experimental
value of 8.4028(4) \AA. The calculated internal $x$ parameter of O was 0.259(4)
\AA\/ (experiment = 0.2604(2)). The very close correspondence in this system
provides confidence for the optimized LiCr$_2$O$_4$ structure. A magnetic
moment of 1.84 $\mu_{\rm B}$ per V was obtained from LMTO calculations on the 
cubic phase (1.85$\mu_{rm B}$ from LAPW), indicating some reduction of the moment from
the spin-only value of 2 $\mu_{\rm B}$. The DOS for V $d$ states and O $p$ 
states for spin-polarized cubic ZnV$_2$O$_4$ is displayed in Fig.~7(a).
Key differences in the DOS of ZnV$_2$O$_4$ and LiCr$_2$O$_4$ include reduced 
metal-oxygen covalency corresponding to fewer V $d$ states in the region 
of O $p$ states. In addition, the Fermi energy no longer lies on a sharp DOS
peak, despite similarities in the shapes of the $t_{2g}(\uparrow)$ manifolds.
Cubic ZnV$_2$O$_4$ is a poorer half-metal when compared with LiCr$_2$O$_4$
as already manifest in the reduced magnetic moment. The $d$ density of
states of one
 of 
the V atoms in antiferromagnetic ZnV$_2$O$_4$ is displayed in Fig.\ \ref{fig07}(b).
The LMTO calculations were performed on the experimental structure 
explicitly considering four independent V atoms forming a distorted 
tetrahedron with a short and a long V-V distance. The antiferromagnetic
structure corresponds to spins on the tetrahedron being antiparallel when 
proximal (2.962 \AA) and parallel when far (2.963 \AA) in correspondence
with experiment \cite{reehuis}. We observe that the DOS remains metallic, 
indicating that correlation must play an important role in this compound, 
and that the one-electron description does not suffice. The changes in the 
DOS on going from the ferromagnetic cubic structure to the antiferromagnetic
tetragonal structure include a slight narrowing of $t_{2g}$ states and
a reduction in the contribution from the DOS peak just below the $E_{\rm F}$.

\subsection{Magnetism of LiCr$_2$O$_4$}

Spin polarized LMTO calculations suggested that the magnetic ground state for 
LiCr$_2$O$_4$ is more stable than the non-magnetic ground state obtained
from a non-spin polarized calculation, by 0.4 eV \textit{per\/} Cr atom.
The converged magnetic moment on Cr was 2.49 $\mu_{\rm B}$ in calculations that
made use of the non-local Perdew-Wang exchange correlation prescription. 
Since there are 2.5 $d$ electrons \textit{per\/} Cr$^{3.5+}$ in LiCr$_2$O$_4$, 
the value of the magnetic moment corresponds to nearly 100\% spin
polarization (an extensive discussion was given in section II B.  

\begin{table}[thb]
\begin{center}
\begin{tabular}{c||c|cccccc}
$a \backslash \delta$ & & -0.02 & -0.01 & 0.00 & 0.01 & 0.015 & 0.02 \\
\hline
\hline
8.00 & $M$ & 0.00 & 0.00 & 2.46 & 2.47 & 2.46 & 2.43 \\
     & $P$ & 0    & 0    &   94 &   90 &   85 &   80 \\
8.05 & $M$ & 0.00 & 2.33 & 2.48 & 2.50 & 2.48 & 2.26 \\
     & $P$ & 0    &   65 &   96 &   91 &   87 &   86 \\
8.10 & $M$ & 0.00 & 2.42 & 2.49 & 2.50 & 2.49 & 2.27 \\
     & $P$ & 0    &   76 &   97 &  100 &   88 &   88 \\
8.15 & $M$ & 0.00 & 2.46 & 2.50 & 2.50 & 2.50 & 2.27 \\
     & $P$ &    0 &   88 &   99 &  100 &   93 &   91 \\
8.20 & $M$ & 0.62 & 2.49 & 2.50 & 2.50 & 2.50 & 2.50 \\
     & $P$ &    9 &   93 &  100 &  100 &  100 &   92 \\
8.25 & $M$ & 2.42 & 2.50 & 2.50 & 2.50 & 2.50 & 2.28 \\
     & $P$ &   83 &  100 &  100 &  100 &  100 &   94 \\
\end{tabular}
\end{center}
\caption{Magnetic properties of LiCr$_2$O$_4$ (LMTO/GGA) as a function of the 
crystal structure. $a$ is the cell parameter in \AA, $\delta$ is the deviation
of the oxygen position from 0.25, $M$ is the magnetization \textit{per\/} Cr
in $\mu_{\rm B}$, and $P$ is the percentage spin polarization of conduction
electrons.} 
\end{table}

Withing the usual von-Barth-Hedin LSDA \cite{vonBarth}, the value was reduced 
to 2.42 $\mu_{\rm B}$ per Cr.  We were interested in how changes in the crystal 
structure might affect the magnetism and half-metallicity in LiCr$_2$O$_4$. 
To this end, we have mapped in Table~1, the magnetization $M$ \textit{per\/} 
Cr, as well as the extent of half-metallicity at $E_{\rm F}$ defined

\begin{displaymath}
P = \frac{N_\uparrow(E_{\rm F}) - N_\downarrow(E_{\rm F})}
            {N_\uparrow(E_{\rm F}) + N_\downarrow(E_{\rm F})} \times 100 \%
\end{displaymath}

\noindent as a function of the two structural parameters, the cubic cell 
parameter $a$ and the structural parameter $\delta$ which is a measure of the
trigonal distortion of CrO$_6$ octahedra as described in section II.
 From Table~1, we see that larger unit cells
favor a larger magnetic moment and larger $P$. This is due to the narrowing
of bands as the separation between atoms becomes larger. Large values of
$\delta$ seem to be contraindicated for magnetism and half-metallicity; the
best half-metals correspond to nearly perfect CrO$_6$ octahedra.
These results were confirmed by the structure optimization procedure of the 
LAPW calculations.

\acknowledgements{ 
This work was partially supported by the MRL program of 
the National Science Foundation under the Award No. DMR00-80034. 
RS would like to acknowledge a start-up grant from the Dean, College of 
Engineering, UCSB. RV thanks A. Krimmel for discussions and
acknowledges financial support from the German Science Foundation.}

\clearpage


\begin{thebibliography}{99}
\bibitem{spinel_struc} Space group $Fd\overline{3}m$: 
A atoms at $(\frac{1}{8}, \frac{1}{8}, \frac{1}{8})$; 
B atoms at $(\frac{1}{2}, \frac{1}{2}, \frac{1}{2})$; 
O atoms at $(x,x,x)$. Alternate choices of the origin are possible in this 
space group.
\bibitem{comment_GGA}
We also performed  LAPW calculations using the Perdew-Wang prescription
without observing for the DOS and bandstructures any appreciable differences 
to the Perdew-Burke-Ernzerhof results.
\bibitem{comment_2}
 For LiV$_2$O$_4$, LiMn$_2$O$_4$ and ZnV$_2$O$_4$ we took respectively
 as energy threshold -7, -7 and -10 Ryd.
\end{thebibliography}
\end{document}